# Automatic Internal Stray Light Calibration of AMCW Coaxial Scanning LiDAR Using GMM and PSO

Sung-Hyun Lee, Yoon-Seop Lim, Wook-Hyeon Kwon, Yong-Hwa Park

*Abstract*— **In this paper, an automatic calibration algorithm is proposed to reduce the depth error caused by internal stray light in amplitude-modulated continuous wave (AMCW) coaxial scanning light detection and ranging (LiDAR). Assuming that the internal stray light generated in the process of emitting laser is static, the amplitude and phase delay of internal stray light are estimated using the Gaussian mixture model (GMM) and particle swarm optimization (PSO). Specifically, the pixel positions in a raw signal amplitude map of calibration checkboard are segmented by GMM with two clusters considering the dark and bright image pattern. The loss function is then defined as L1-norm of difference between mean depths of two amplitude-segmented clusters. To avoid overfitting at a specific distance in PSO process, the calibration check board is actually measured at multiple distances and the average of corresponding L1 loss functions is chosen as the actual loss. Such loss is minimized by PSO to find the two optimal target parameters: the amplitude and phase delay of internal stray light. According to the validation of the proposed algorithm, the original loss is reduced from tens of centimeters to 3.2 mm when the measured distances of the calibration checkboard are between 1 m and 4 m. This accurate calibration performance is also maintained in geometrically complex measured scene. The proposed internal stray light calibration algorithm in this paper can be used for any type of AMCW coaxial scanning LiDAR regardless of its optical characteristics.**

## I. INTRODUCTION

LIGHT detection and ranging (LiDAR) technology has enormously improved the 3D recognition performance of various intelligent systems such as drones, autonomous vehicles, and robots [1], [2]. Compared to conventional stereo vision, which takes relatively long time for disparity calculation, the LiDAR can provide precise 3D depth information in real time with relatively low calculation loads. In addition to the highly precise measurement performance, the relatively low cost of LiDAR compared to structured light (SL)-based active illumination method also increases the versatility of LiDAR in various engineering applications [3], [4]. Based on these advantages of LiDAR, many researchers have developed the LiDAR systems and related recognition technologies.

Among LiDAR systems, there exist two main methods used to measure the distance: direct time-of-flight (ToF) method and indirect ToF method. The direct ToF measurement method utilizes highly precise time-to-digital converter (TDC) to directly measure the travel time of emitted light signals [5]. Due to the simplicity of the measurement principle, the post processing algorithm for direct ToF LiDAR sensors is generally simpler than that of indirect ToF sensors. Meanwhile, the indirect ToF method also referred to amplitude-modulated continuous wave (AMCW) method, estimates the phase delay of light signals reflected from an object using signal demodulation [6], [7]. The AMCW ToF sensor is widely used for relatively short ranges, up to 10 m, due to its high measurement precision and low cost compared to the direct ToF LiDAR sensor. For every application situations, there is an appropriate distance measurement method in the aspects of maximum range, object property, frame rate, etc.

Aforementioned LiDAR systems have a common measurement error source, the internal stray light. Although the optical components in LiDAR systems such as beam splitter (BS) and focusing lens are coated by anti-reflection materials, 100 % penetration or reflection of light is impossible. Namely, unwanted scattered or multi-reflected light inevitably exists inside the optical lenses of LiDAR systems which results in depth distortion [5], [8], [9]. Internal stray light is mainly affected by the structure of the optical components in LiDAR system, *i.e.*, the relative orientation/position of each lens.

Considering the intrinsic model of LiDAR, many researchers have developed internal stray light mitigation methods [5], [8]-[11]. For the scanning type LiDAR, many researchers have used single ultra-high precision TDC or multi-channel of TDC to directly estimate the parameters of internal stray light, *i.e.*, the time delay and amplitude of stray light [5], [10]. Since TDC directly records the arrival time of light in real-time, it is possible to separate the original reflected light from multiple stray lights. However, this TDC-based approach is generally associated with high costs. Although there exists a signal processing-based approach to estimate the scattered light, this method is not applicable in real-time fast imaging due to its complex feedback algorithm structure [11]. For the flash type ToF sensors (ToF cameras), there are two main approaches to estimate stray light information: heterodyne modulation and hardware optimization of the optical layout. The heterodyne modulation method utilizes multiple modulation frequencies to increase the information of acquired cross-correlations including the stray light [9]. After post-processing the data acquired by heterodyne mixing, the undistorted distance can be estimated. The coded modulation method, similar to amplitude modulation with multiple frequencies, has also been used to estimate the stray light information in previous research works [9]. For the hardware optimization, some researchers have tried

Sung-Hyun Lee, Yoon-Seop Lim, and Yong-Hwa Park are with the Department of Mechanical Engineering, Korea Advanced Institute of Science and Technology, Republic of Korea. (*Corresponding author*: Yong-Hwa Park.)

Wook-Hyeon Kwon is with the Mechatronics Research, Samsung Electronics Co., Republic of Korea.

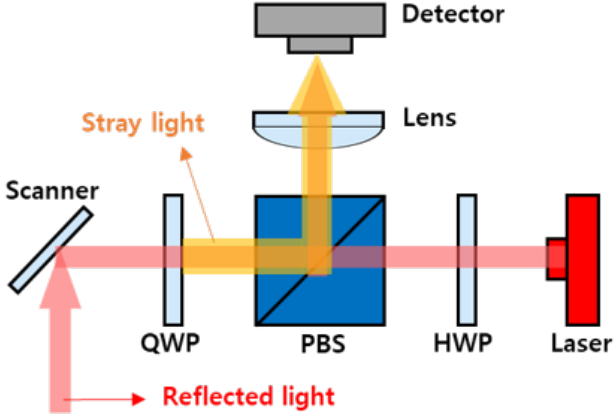

Fig. 1. Example of internal stray light in homodyne AMCW LiDAR optical system. QWP is quarter wave plate, PBS is polarizing beam splitter, HWP is half wave plate.

to modify the layout of optical components using optical path simulation to minimize the stray light effects in ToF sensors [8]. Likewise, for flash-type ToF sensors, there exist stray light mitigation methods mainly related to the modification of modulation source or optical layout. These hardware modifications inevitably increase the cost of sensors.

To mitigate internal stray light effects without aforementioned high-cost hardware modifications, this paper proposes an automatic internal stray light calibration method targeting coaxial scanning type AMCW LiDAR [6]. Due to the existence of static-internal stray light generated in the process of emitting laser, the ratio of directly reflected light from an object to the internal stray light varies with the reflectivity of the object even at same distance. Consequently, the measured distance is inevitably changed with the amplitude of received light even at same object distance. If the exact stray light information can be estimated, then the depth distortion caused by stray light can be mitigated [12]. To precisely estimate the amplitude and phase delay of internal stray light using a single modulation frequency, the Gaussian mixture model (GMM) and particle swarm optimization (PSO) are used in this paper [13], [14]. Specifically, a calibration checkboard is measured with previously developed AMCW coaxial scanning LiDAR at multiple fixed distances [6]. The image pixel positions in each raw amplitude (amplitude of cross-correlation) map are then segmented into two clusters: a pixel position group of bright pattern, and that of dark pattern. This amplitude-based segmentation is processed using GMM, since the data distributions of measured depth and raw amplitude maps follow the Gaussian distribution [15]. After such amplitude-based segmentation, the L1-norm of difference between the mean depths of corresponding two amplitude-segmented clusters is calculated for each measured distance case. The average of all these aforementioned L1-norms is then chosen as actual loss to be minimized by PSO. By the optimization process of PSO, the optimal values of the two target parameters, *i.e.*, the amplitude and phase delay of internal stray light, are extracted. Using these estimated stray light parameters, the cross-correlations caused by stray light are calculated and subtracted from the raw measured cross-correlations to result in corrected cross-correlations. Using these corrected cross-correlations, post-corrected depth maps can then be generated. Consequently, this optimization process is the same as finding the internal stray light parameters which make all post-corrected depth maps of the checkboard as flat as possible. After finding out the correct stray light parameters, these values can also be used to correct depth maps of other object scenes. Experimental validation in this paper showed that there was a decrease in loss from tens of centimeters to 3.2 mm when the distance of the calibration checkboard ranged from 1 m to 4 m. Such highly precise depth error correction is also maintained in complex multi-objects images. The main advantages of the proposed internal stray light calibration method can be summarized as follows:

    1) As there is no systematic assumption for stray light parameter identification, the proposed calibration method can be utilized in any type of AMCW coaxial scanning LiDAR.

    2) The proposed calibration method utilizes only some depth and raw amplitude maps of checkboard with single modulation frequency.

This paper is organized as follows: Section II presents the problem statement related to internal stray light. Section III presents the internal stray light calibration method using GMM and PSO. Section IV presents the validation results of the proposed stray light calibration method including parametric study and experimental results. Section V presents the conclusion of this paper.

## II. PROBLEM DEFINITION: INTERNAL STRAY LIGHT IN AMCW COAXIAL SCANNING LIDAR

In AMCW coaxial scanning LiDAR, the light source, which is generally a laser diode, is amplitude-modulated in sinusoidal waveform [6], [16]. The modulated light signal is then collimated and emitted to the measurement point through optical components such as BS, wave plate, and scanner. After the emitted light signal is reflected from the object, it is focused on the active area of the photodetector, such as the avalanche photodiode (APD). Using the received light signal and the demodulation (reference) signal, the cross-correlation samples are calculated. Based on these correlation samples, the phase, amplitude, and offset of the original cross-correlation function can be estimated [6], [16]. However, in the process of emitting laser signal, unintended light signals are generated by inner multi-reflection in optics and sensed by APD. Such internal stray light results in distortion of cross-correlation as shown in Fig. 1.

In Fig. 1, there exists multi-reflected light in optics between each reflectance facet during emitting laser signal to object. Although only one stray light signal is presented in Fig. 1 as an example, there actually exist lots of stray light rays which are multi-reflected or scattered inside the coaxial optics during emitting the laser signal. All these internal stray light signals can be assumed as static if the measurement conditions such as layout of optical components, the laser power, and the modulation frequency are fixed. Consequently, for homodyne

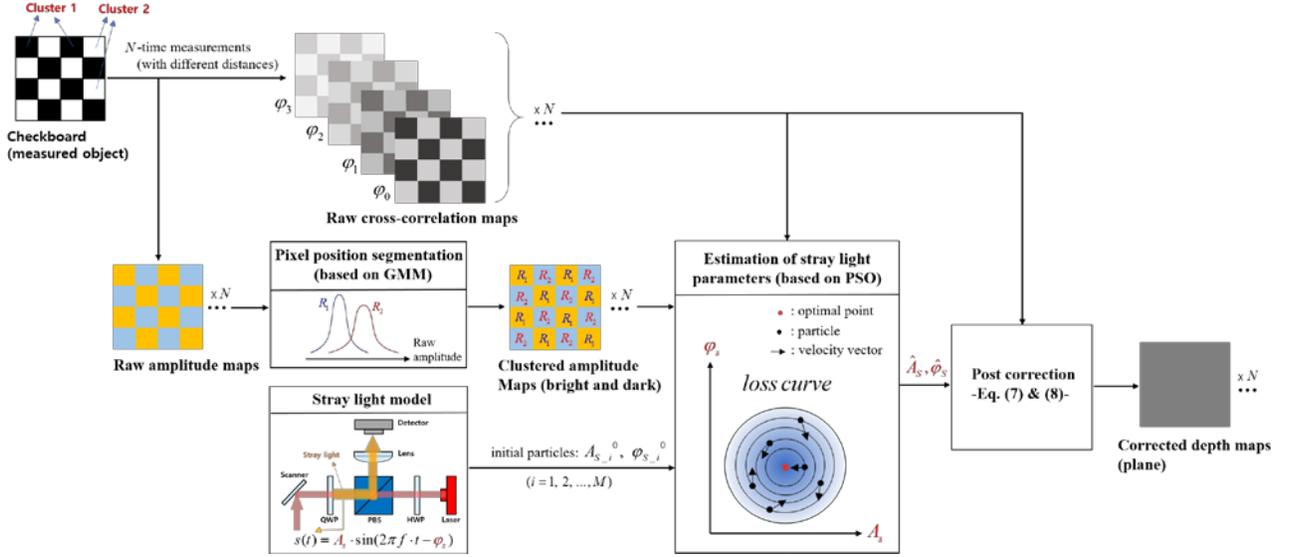

Fig. 2. Block diagram of automatic stray light calibration method using calibration checkboard. $A_s$ is the amplitude of internal stray light, $\varphi_s$ is the phase delay of internal stray light, $\hat{A}_s$ is the estimated amplitude of internal stray light by PSO, $\hat{\varphi}_s$ is the estimated phase delay of internal stray light by PSO, $N$ is the number of measured distances, $h$ is the number of particles, and $m_{\varphi_n}(t)$ is the demodulation signal defined in (4). For one measurement at specific distance, 4 cross-correlation sample maps ( $\varphi_n = \frac{n\pi}{2}$ ($n = 0,1,2,3$) ) and one raw amplitude map are acquired.

mixing which modulates signal in sinusoidal waveform with single frequency, the net static internal stray light signal can be modeled as a single sinusoidal waveform following trigonometric characteristics [12]. Based on this fact, the related mathematical expressions in the time domain are as follows:

$$\hat{C}(\varphi_n) = \frac{1}{T_{int}} \int_0^{T_{int}} \left( r(t) + s(t) \right) \cdot m_{\varphi_n}(t)\,dt$$

$$= \frac{1}{T_{int}} \int_0^{T_{int}} r(t) \cdot m_{\varphi_n}(t)\,dt + \frac{1}{T_{int}} \int_0^{T_{int}} s(t) \cdot m_{\varphi_n}(t)\,dt$$

$$= C_r(\varphi_n) + C_s(\varphi_n) \tag{1}$$

$$r(t) = A_r \cdot \sin(2\pi f \cdot t - \varphi_r) + B_r, \tag{2}$$

$$s(t) = A_s \cdot \sin(2\pi f \cdot t - \varphi_s) + B_s \tag{3}$$

$$m_{\varphi_n}(t) = m \cdot \sin(2\pi f \cdot t + \varphi_n) \tag{4}$$

$$\hat{\Gamma} = \frac{\sqrt{\left(\hat{C}(\varphi_3) - \hat{C}(\varphi_1)\right)^2 + \left(\hat{C}(\varphi_2) - \hat{C}(\varphi_0)\right)^2}}{2} \tag{5}$$

$$\hat{\varphi} = \arctan\left( \frac{\hat{C}(\varphi_3) - \hat{C}(\varphi_1)}{\hat{C}(\varphi_0) - \hat{C}(\varphi_2)} \right) \tag{6}$$

where $\hat{C}(\varphi_n)$ is raw measured cross-correlation sample, $C_r(\varphi_n)$ is cross-correlation sample generated by directly reflected light signal, $C_s(\varphi_n)$ is cross-correlation generated by stray light signal, $T_{int}$ is integration time, $r(t)$ is directly reflected light signal, $A_r$ is amplitude of directly reflected light signal, $B_r$ is offset of directly received light signal, $\varphi_r$ is phase delay of directly reflected light signal, $s(t)$ is internal stray light signal, $A_s$ is amplitude of stray light signal, $B_s$ is offset of stray light signal, $\varphi_s$ is phase delay of stray light signal, $m_{\varphi_n}(t)$ is phase-shifted demodulation signal, $\varphi_n$ is $n^{th}$ phase shift, $m$ is amplitude of demodulation signal which is 0.4785 in voltage in this paper, $\hat{\Gamma}$ is the amplitude of raw cross-correlation, and $\hat{\varphi}$ is the measured phase delay of raw cross-correlation. As shown in (1), the net recorded cross-correlation is the linear summation of cross-correlations generated by directly reflected light and stray light. In general, $n$ is fixed as 4, i.e., $\varphi_n = \frac{n\pi}{2}$ ($n = 0,1,2,3$) [6], [17]. Meanwhile, all the DC terms in (2) and (3) are negligible since all DC terms are deleted in (6).

If the parameters of internal stray light in (3), i.e., $A_s$ and $\varphi_s$, are identified, the exact $C_s(\varphi_n)$ can be estimated. By subtracting estimated $C_s(\varphi_n)$ from actually measured correlation sample, $\hat{C}(\varphi_n)$, the corrected correlation sample $C_r(\varphi_n)$ is extracted. Using the corrected correlation samples, the true depth corresponding to $\varphi_r$ can be estimated as follow:

$$C_s(\varphi_n) = \frac{1}{T_{int}} \int_0^{T_{int}} s(t) \cdot m_{\varphi_n}(t)\,dt$$

$$= \frac{1}{T_{int}} \int_0^{T_{int}} \left\{ A_s \cdot \sin(2\pi f \cdot t - \varphi_s) \right\} \cdot m_{\varphi_n}(t)\,dt \tag{7}$$

$$\tilde{d} = \frac{c}{4\pi \cdot f} \cdot \arctan\left( \frac{\left(\hat{C}(\varphi_3) - C_s(\varphi_3)\right) - \left(\hat{C}(\varphi_1) - C_s(\varphi_1)\right)}{\left(\hat{C}(\varphi_0) - C_s(\varphi_0)\right) - \left(\hat{C}(\varphi_2) - C_s(\varphi_2)\right)} \right)$$

$$= \frac{c}{4\pi \cdot f} \cdot \arctan\left(\frac{C_r(\varphi_3) - C_r(\varphi_1)}{C_r(\varphi_0) - C_r(\varphi_2)}\right) \quad (8)$$

where $\tilde{d}$ is the corrected depth , *i.e.*, estimated true depth corresponding to $\varphi_r$, based on the identified parameters of internal stray light. To acquire the correct depth corresponding to directly reflected light in (8), the accurate estimation of internal stray light parameters, *i.e.*, $A_s$ and $\varphi_s$, is important.

In this paper, to estimate the aforementioned parameters of net internal stray light and to correct the distorted correlation samples, a novel calibration method based on GMM and PSO is proposed. The key idea of the proposed method is that the signal-to-noise ratio (SNR) of directly reflected light to static internal stray light changes with the reflectivity of an object even at same distance. This different SNR directly results in the different measured distances. Following this measurement property, the raw measured depth map of the calibration checkboard has two distinguished depth clusters due to the repetitive dark and bright image pattern. By finding the target parameters of stray light that make the post-corrected depth map of calibration checkboard flat, the true depth induced by only reflected light can be properly estimated using (8). The detailed explanation of the stray light calibration method is presented in the following Section.

## III. INTERNAL STRAY LIGHT CALIBRATION IN AMCW COAXIAL SCANNING LIDAR USING GMM AND PSO

In this Section, the details of the stray light calibration method are presented. The block diagram of the proposed calibration method is shown in Fig. 2. According to Fig. 2, 4 cross-correlation sample maps and one raw amplitude (amplitude of cross-correlation) map of checkboard are measured at $N$−different object distances. The GMM-based segmentation of image pixel positions in a raw amplitude map is then processed $N$− times to classify two different reflectivity regions (dark and bright) for each raw amplitude image. The reason GMM is used in this paper is mainly attributed to the inherent Poisson distributions of measured depth and raw amplitude data in this paper. As the number of data points is over 500, this Poisson distribution can be approximated as a Gaussian distribution [15], [17]. Namely, it can be intuitively deduced that the raw amplitude data distribution of the checkboard measured by AMCW coaxial scanning LiDAR is composed of two Gaussian clusters corresponding to the dark and bright repetitive image pattern. After the amplitude-based segmentation of image pixel positions, the loss at specific measured distance case is defined as L1 norm of difference between mean depths of corresponding two pixel position clusters. Likewise, $N$−different losses are calculated for each measured distance case. The average of these L1 losses is then chosen as actual loss. This loss is minimized by PSO algorithm, which is a non-convex iterative optimization method [13], [20]. Considering multiple local minimums of searching points and global minimum simultaneously, the PSO algorithm can find the exact global optimal point in relatively fast calculation time. The estimated stray light parameters are then used to correct the cross-correlation samples. At final step, the post-corrected

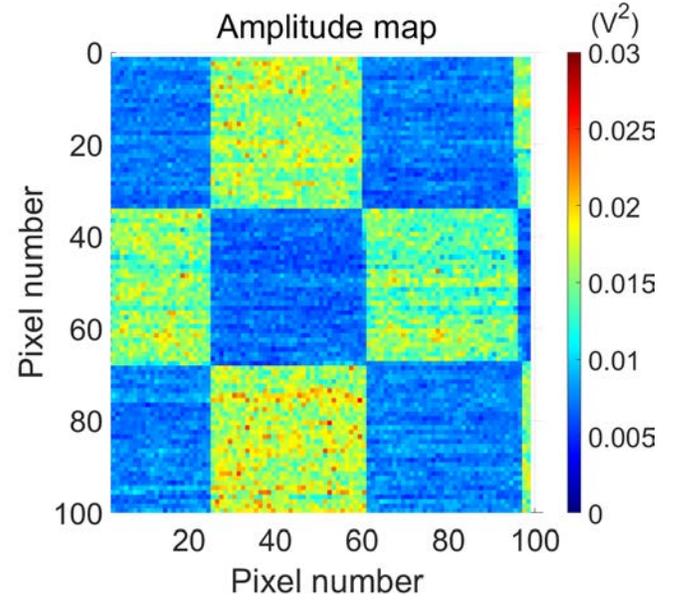

Fig. 3. Raw amplitude map of calibration checkboard measured by AMCW scanning LiDAR at distance of 2.3 m.

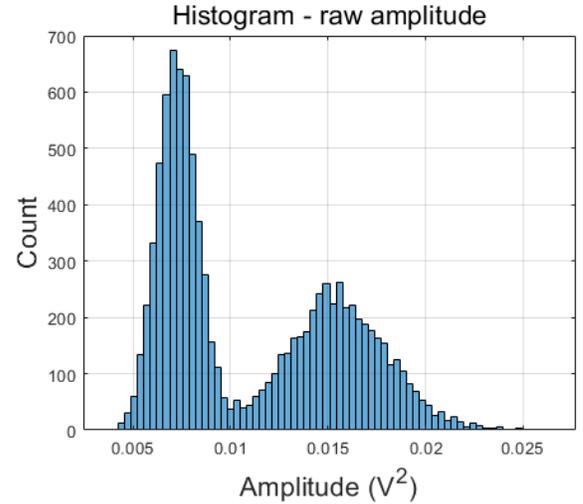

Fig. 4. Histogram of raw amplitude map measured by AMCW scanning LiDAR at distance of 2.3 m

depth maps are generated based on (8).

To specifically explain the backgrounds of Fig. 2, the GMM, PSO, and depth correction method are presented.

### A. Gaussian Mixture Model (GMM)

GMM is generally used to cluster the given dataset assuming that the distribution of data follows a linear combination of multiple Gaussian distribution functions [14], [18], [19]. If the number of clusters is fixed in advance, the GMM can classify each data into one of the Gaussian distributed clusters. To determine the class of a given input data, the responsibility, *i.e.,* the posterior probability of belonging to specific cluster for the input data, is calculated for each cluster. According to GMM principle, the input data is determined to belong to the class of which the responsibility is the maximum [14], [18], [19]. The detailed expression for the pixel position segmentation based on GMM in Fig. 2 is expressed as follow [14], [18], [19]:

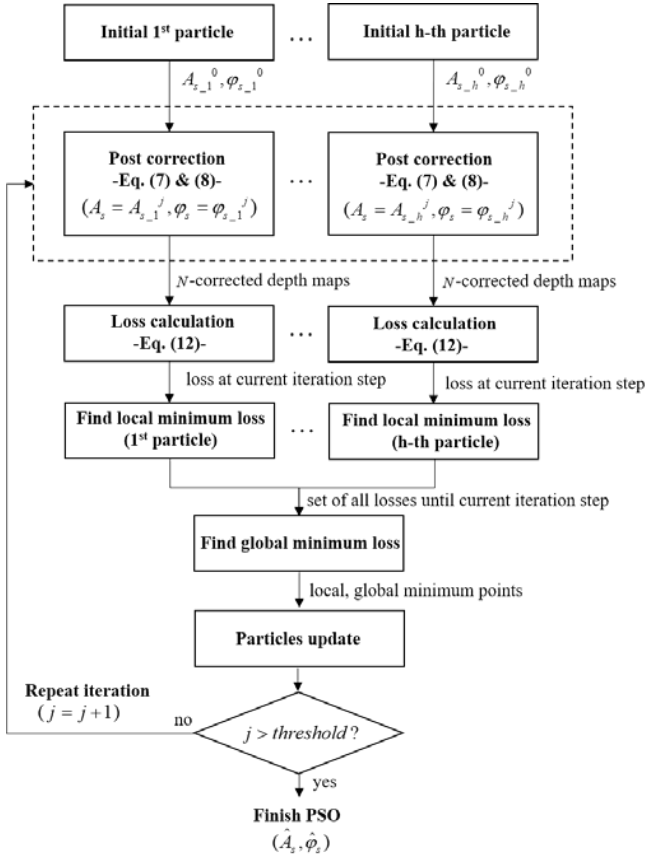

Fig. 5. Flow architecture representing estimation of stray light parameters based on PSO in Fig. 2. $h$ is the number of particles. $N$ is the number of measured distances in Fig. 2.

$$\gamma_c^{\ k}\left(\hat{\Gamma}_k(u,v)\right) = \frac{\pi_c^{\ k} \cdot \mathrm{N}(\hat{\Gamma}_k(u,v) \mid \mu_c^{\ k}, \sigma_c^{\ k})}{\sum_{l=1}^{2} \pi_l^{\ k} \cdot \mathrm{N}(\hat{\Gamma}_k(u,v) \mid \mu_l^{\ k}, \sigma_l^{\ k})} \qquad (9)$$

where $\hat{\Gamma}_k(u,v)$ is the raw amplitude of cross correlation at the pixel position of $(u,v)$ in the $k^{th}$ ($k = 1, 2, \ldots N$) raw amplitude map, $\gamma_c^{\ k}(\bullet)$ is the responsibility corresponding to cluster $c$ (1: dark pattern, 2: bright pattern) for input data in the $k^{th}$ raw amplitude map, $\pi_c^{\ k}$ is the weight factor of cluster $c$ in the $k^{th}$ raw amplitude map, $\mu_c^{\ k}$ is the mean-amplitude of image pixels belonging to cluster $c$ in the $k^{th}$ raw amplitude map, $\sigma_c^{\ k}$ is the variance of raw amplitude of image pixels belonging to cluster $c$ in the $k^{th}$ raw amplitude map, and $\mathrm{N}(\mu_c^{\ k}, \sigma_c^{\ k})$ is the Gaussian distribution function of raw amplitude corresponding to class $c$ in the $k^{th}$ raw amplitude map. For each $k^{th}$ ($k = 1, 2, \ldots N$) raw amplitude map, all image pixels are determined to belong to either pixel position cluster of dark pattern, i.e., $R_1^{\ k}$, or pixel position cluster of bright pattern, i.e., $R_2^{\ k}$, considering the maximum value of (9). Meanwhile, to estimate the proper parameters, i.e., $\pi_c^{\ k}$, $\mu_c^{\ k}$, $\sigma_c^{\ k}$ for all $c$ and $k$, Expectation-Maximization (EM) is used in this paper like many other previous works [14], [18], [19].

As presented above, the GMM assumes the distribution of dataset same as the linear summation of Gaussian clusters. Due to this assumption, the GMM is not always the ideal solution for various clustering and segmentation problems. However, for the data distribution measured by AMCW LiDAR, the GMM intuitively fits well. Since the inherent distribution of large number data measured by AMCW LiDAR definitely follows the Gaussian distribution [6], [15], [17], the raw amplitude map of calibration checkboard can be divided into two main Gaussian clusters, i.e., dark pattern and bright pattern. To validate such tendency, Fig. 3 and 4 are presented. According to Fig. 3, the raw amplitude map of checkboard at distance of 2.3 m has clear separation for each reflectivity region. The histogram of corresponding amplitude data in Fig. 4 also shows the two main Gaussian-like clusters. Considering these backgrounds, the GMM was utilized in this paper to segment the raw amplitude map of calibration checkboard.

### B. Particle Swarm Optimization (PSO)

PSO is an iterative optimization method mimicking the natural movement mechanism of a flock of birds or a school of fish [13], [20]. The PSO randomly sets a number of searching points, which are also called *particles*, in the data feature space. Each particle moves in the feature space following the direction of minimizing the designed loss. The direction is directly affected by the velocity vector in current optimization iteration step. To update the particles for each iteration, the velocity vectors are properly updated. The distinguishing point is that the velocity vector is updated considering not only the local minimum found by specific particle, but also the global minimum found by the entire particles in the current iteration step. This indicates that each particle shares its current feature space information at every iteration step, which resembles the nature social behavior of animals. Such local-global iterative optimization has robust convergence performance even for non-convex optimization problems such as complex calibration problems of LiDAR and cameras [21]–[24]. Such PSO is adopted as optimization method in this paper to estimate the internal stray light parameters, i.e., $\hat{A}_s$ and $\hat{\varphi}_s$ in Fig. 2. The overall block diagram of PSO process is shown as Fig. 5. As shown in Fig. 5, total $h$ − candidates of stray light parameter vector exist. For each candidate (particle) assumption, post-correction based on (7) and (8) are processed for $N$ − different measured distance cases in Fig. 2. Consequently, $N$ − corrected depth maps are generated for each particle. Based on the $N$ − corrected depth maps, the local loss is calculated for each particle. After finding local-global minimums and particle updates, the aforementioned process is repeated until the total PSO iteration number satisfies predetermined threshold.

For the precise estimation of internal stray light in Fig. 5, the loss should be properly defined to reduce the mean depth discrepancy between $R_1^{\ k}$ and $R_2^{\ k}$ in corresponding $k^{th}$ ($k = 1, 2, \ldots N$) raw amplitude map, considering all cases of $k$. For the specific particle, the loss and related parameters in Fig. 2 and 5 are defined as follows:

$$x_i^{\ j} = \left(A_{s\_i}^{\ j}, \varphi_{s\_i}^{\ j}\right)^T \qquad (10)$$

TABLE I
Hyperparameters of GMM

| Hyperparameter | Value |
|---|---|
| Maximum iteration of EM | 1,000 |
| Termination tolerance for log-likelihood function | $1.0 \cdot 10^{-6}$ |
| Confidence margin of posterior probability | 0.9 |

TABLE II
Hyperparameters of PSO

| Hyperparameter | Value |
|---|---|
| Range of weight decay for velocity vector | [0.1, 1.1] |
| Weight factor of local minimum | 1.49 |
| Weight factor of global minimum | 1.49 |
| Number of particles ($h$ in Fig. 5) | 20 |
| Maximum iteration (threshold in Fig. 5) | 100 |
| Tolerance of loss variation | $1.0 \cdot 10^{-6}$ |
| Stall iteration number | 20 |

$$\bar{d}_i^{\,j}(R_c^{\,k}) = \frac{1}{n(R_c^{\,k})} \sum_{\forall (u,v) \in R_c^{\,k}} \tilde{d}_i^{\,j}(u,v,k) \quad (11)$$

$$L_i^{\,j} = \frac{1}{N} \cdot \sum_{k=1}^{N} \left\| \bar{d}_i^{\,j}(R_1^{\,k}) - \bar{d}_i^{\,j}(R_2^{\,k}) \right\| \quad (12)$$

where $x_i^{\,j}$ is the position vector of $i^{th}$ ($i = 1, 2, ... h$) particle at PSO iteration step $j$, $A_{s\_i}^{\,j}$ is the amplitude of internal stray light corresponding to $i^{th}$ particle at PSO iteration step $j$, $\varphi_{s\_i}^{\,j}$ is the phase delay of internal stray light corresponding to $i^{th}$ particle at PSO iteration step $j$, $R_c^{\,k}$ is the pixel position cluster $c$ (1 for dark pattern, 2 for bright pattern) in $k^{th}$ ($k = 1, 2, ... N$) raw amplitude map in Fig. 2, $\tilde{d}_i^{\,j}(u,v,k)$ is the depth value at pixel position $(u,v)$ in $k^{th}$ depth map corrected by post-correction in Fig. 5 utilizing $i^{th}$ particle at PSO iteration step $j$, $n(R_c^{\,k})$ is the total number of pixels belonging to cluster $R_c^{\,k}$, $\bar{d}_i^{\,j}(R_c^{\,k})$ is the mean depth of image pixels belonging to cluster $R_c^{\,k}$ in $k^{th}$ depth map corrected by post-correction in Fig. 5 utilizing $i^{th}$ particle at PSO iteration step $j$, and $L_i^{\,j}$ is averaged L1 loss based on $i^{th}$ particle at PSO iteration step $j$. The physical meaning of (12) is same as the average of the absolute difference between mean depths of pixel position clusters in corresponding corrected depth map based on $i^{th}$ particle at PSO iteration step $j$. The reason to average the L1 loss using multiple calibration checkboard images in (12) is to avoid overfitting at specific measured distance in optimization process. Minimizing the loss in (12) makes it feasible to find the optimal internal stray light parameters that make the corrected depth maps of the calibration checkboard flat.

Based on the aforementioned defined particle and loss, simplified mathematical expressions of PSO are expressed as follows:

$$v_i^{\,j+1} = w^j \cdot v_i^{\,j} + c_1 \cdot r_1 \cdot \left( p_i^{\,j} - x_i^{\,j} \right) + c_2 \cdot r_2 \cdot \left( g^j - x_i^{\,j} \right) \quad (13)$$

$$x_i^{\,j+1} = x_i^{\,j} + v_i^{\,j+1} \quad (14)$$

where $v_i^{\,j}$ is the velocity vector of $i^{th}$ particle at iteration step $j$, $w^j$ is the weight decay which is exponential function of 0.5, $p_i^{\,j}$ is the local minimum point found by $i^{th}$ particle until iteration step $j$, $g^j$ is the global minimum point found by entire particles until iteration step $j$, $c_1$ and $c_2$ are the weight factors, $r_1$ and $r_2$ are random numbers between 0 and 1. The PSO was utilized to find global-minimum loss and corresponding parameters of internal stray light as shown in Fig. 2 and 5.

### C. Depth Correction Using Corrected Cross-Correlation

Using the optimal stray light parameters, the cross-correlation samples generated by stray light, i.e., $C_s(\varphi_n)$, can be calculated based on (7). Each correlation sample of stray light is subtracted from the corresponding raw measured cross-correlation sample, i.e., $\hat{C}(\varphi_n)$, to extract the corrected cross-correlation sample. The corrected cross-correlation samples are the same as the cross-correlation samples of directly reflected light signal, i.e., $C_r(\varphi_n)$. The corrected depth map is then acquired by (8) at final step in Fig. 2.

## IV. VALIDATION OF INTERNAL STRAY LIGHT CALIBRATION METHOD

To validate the calibration performance of the proposed algorithm in Fig. 2, the parametric study and experimental validation are presented in this Section. In the parametric study, the calibration performance of the proposed algorithm for a number of different input images is analyzed in terms of convergence speed and loss value. To validate the actual calibration performance, the experimental results are also presented and analyzed. For measurement objects, calibration checkboard and sculptures were used.

### A. Parametric Study

To utilize GMM and PSO, the related hyperparameters should be properly determined before executing the algorithm. The main hyperparameters related to GMM and PSO are shown in Table I and Table II. All details related to the hyperparameters are explained in other previous works [13], [14], [18], [20], [25].

The image data of the calibration checkboard were acquired for four distance points: 1.75 m, 2.3 m, 3.0 m, and 4.0 m. To analyze the convergence speed, minimized loss, and overfitting of the proposed calibration method in Fig. 2, optimization processes were conducted multiple times changing the number of input images. The CPU used in this paper is an Intel(R) core i5-8500 with 16 GB RAM. The software used is Matlab 2022b. The loss plots along the optimization iteration for different number of input images are shown in Fig. 6. The loss in Fig. 6 is the L1 loss in (12) corresponding to the global optimal point

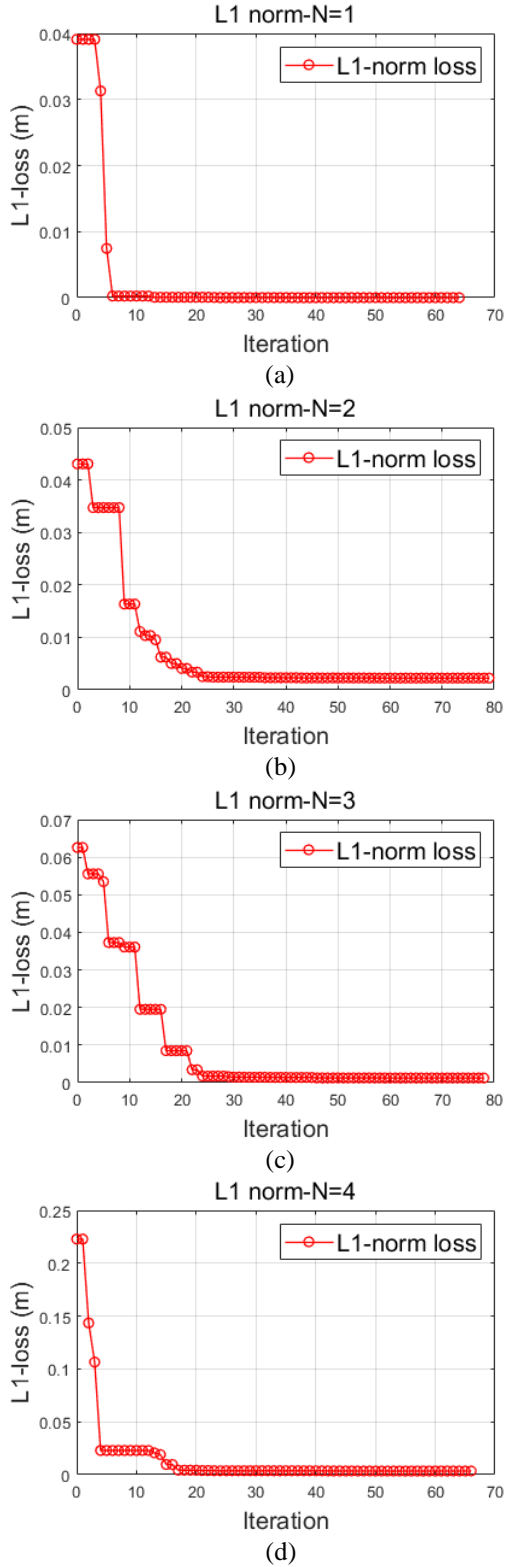

Fig. 6. Loss plot along the iteration for different number of used input images for optimization: (a) distance of input images is 1.75 m, (b) distances of input images are 1.75 m, 4.0 m, (c) distances of input images are 1.75 m, 2.3 m, 4.0 m, (d) distances of input images are 1.75 m, 2.3 m, 3.0 m, 4.0 m.

for each PSO iteration step in (13). According to Fig. 6, the

### TABLE III
### COMPARISON OF L1 LOSS AFTER CORRECTION USING DIFFERENT INPUT IMAGES

| Distances of input images used for estimation of stray light parameters in Fig. 2 (m) | L1 loss calculated using only checkboard images measured at 3.0 m (m) |
|---|---|
| 1.75 | 0.3128 |
| 1.75, 4.0 | 0.1560 |
| 1.75, 2.3, 4.0 | 0.0151 |

### TABLE IV
### COMPARISON OF PROCESSING TIME FOR DIFFERENT INPUT IMAGES

| Distances of input images used for estimation of stray light parameters in Fig. 2 (m) | Total processing time (sec) |
|---|---|
| 1.75 | 28.540 |
| 1.75, 4.0 | 65.218 |
| 1.75, 2.3, 4.0 | 104.613 |
| 1.75, 2.3, 3.0, 4.0 | 172.038 |

smooth convergence of loss minimization is achieved for all cases of optimization trials. From Fig. 6(a) to Fig. 6(d), the minimized losses are $5.2310 \cdot 10^{-9}$ m, 0.0022 m, 0.0012 m and 0.0032 m, respectively. However, there is no confidence to assure that there is no overfitting at specific distance. To examine the overfitting property, an additional parametric study was conducted as shown in Table III. For each case of different input images shown as left column of Table III, corresponding L1 losses in (12) calculated using only checkboard images measured at 3.0 m ( $N = 1$ ) are presented. For each calculation of L1 loss at 3.0 m, the used internal stray light parameters, $i.e.$, $\hat{A}_s$ and $\hat{\varphi}_s$ in Fig. 2, are estimated from the input images measured at corresponding distance condition shown in left side of Table III. According to Table III, there exists a tendency of overfitting confined to the inner region of input distances. Specifically, if the measured distance of used input images is only 1.75 m, the corresponding L1 loss at 3.0 m in Table III is 0.3128 m, which is much larger than the minimized loss in Fig. 6(a) same as $5.2310 \cdot 10^{-9}$ m. Such large discrepancy between loss in Table III and corresponding minimized loss in Fig. 6(a) means that the estimated stray light parameters are over fitted at 1.75 m. To avoid such overfitting at specific distance, enough input images measured at wide range are necessary. For the case of 2 input distances: 1.75 m and 4.0 m, the corresponding L1 loss in Table III is about 0.1560 m, which is not still enough to avoid overfitting. However, if the input images measured at 1.75 m, 2.3 m, and 4.0 m are used, the corresponding L1 loss at 3.0 m is enormously reduced to 0.0151 m. The gap between averaged loss in Fig. 6(c) and corresponding loss at 3.0 m in Table III is about 1 cm. This gap is quite tolerable value in that the depth deviation due to random shot noises in this paper is cm-scale at 3.0 m, which is described in detail in the following subsection. To ensure margin preventing the over fitting, total 4 input images measured at 1.75 m, 2.3 m, 3.0 m, and 4.0 m are used to estimate the parameters of internal stray light in this paper. The averaged loss in Fig. 6(d) is 0.0032 m. Meanwhile, to test the calculation load, the total processing time for each different input images are also presented in Table IV. Table IV shows that total calculation time for optimization is generally proportional to the number of input images. For the case of 4 input images, the total processing time is 172.038 sec, which is quite tolerable. According to the parametric study, the final optimized phase delay of stray light using 4 input images is 0.3509 in radian, and the corresponding optimized amplitude is 0.0976 in voltage, in this paper.

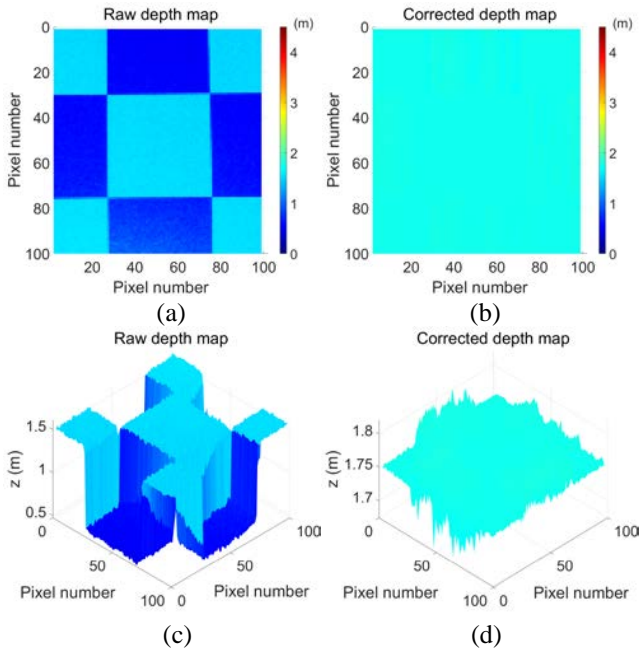

Fig. 7. Depth maps of calibration check board at distance of 1.75 m: (a) raw depth map in front view, (b) corrected depth map in front view, (c) raw depth map in side view, (d) corrected depth map in side view.

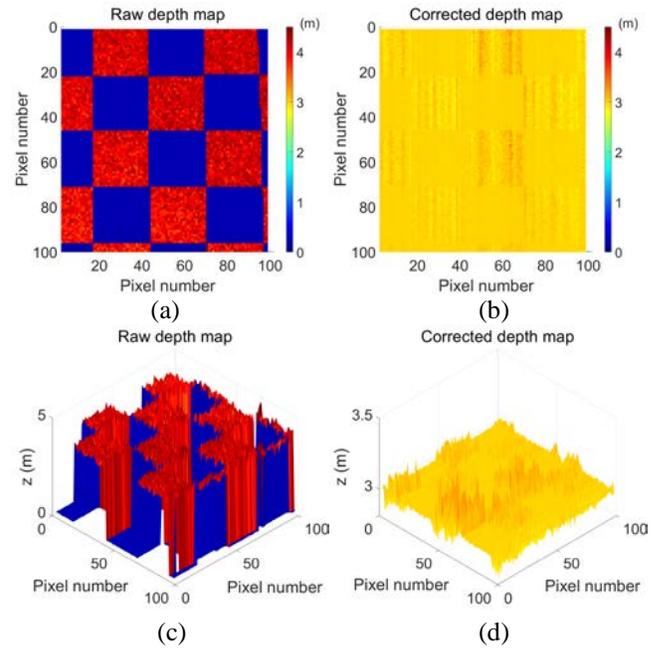

Fig. 9. Depth maps of calibration check board at distance of 3.0 m: (a) raw depth map in front view, (b) corrected depth map in front view, (c) raw depth map in side view, (d) corrected depth map in side view.

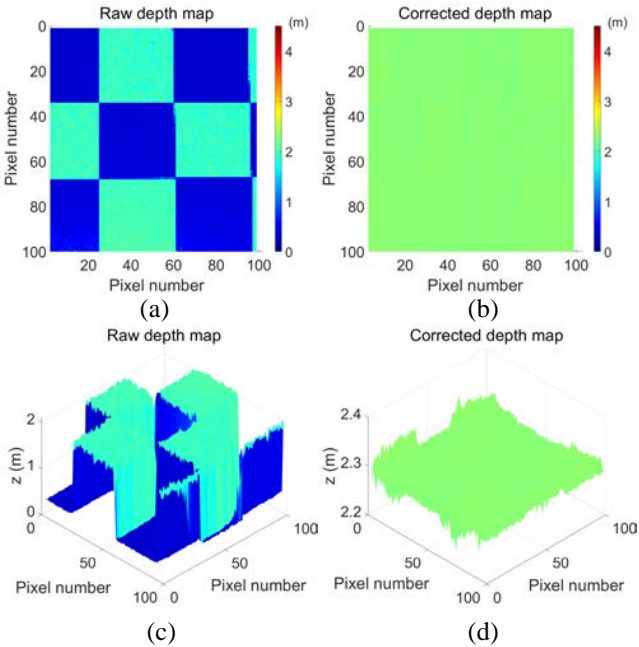

Fig. 8. Depth maps of calibration check board at distance of 2.3 m: (a) raw depth map in front view, (b) corrected depth map in front view, (c) raw depth map in side view, (d) corrected depth map in side view.

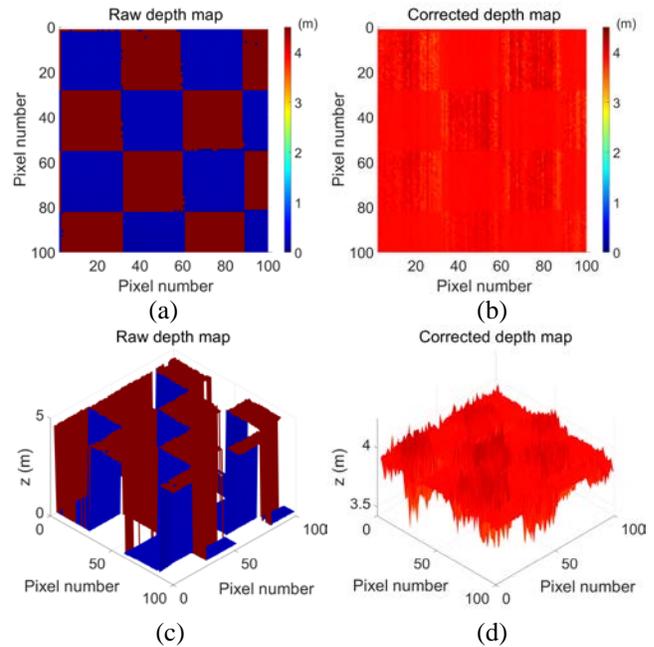

Fig. 10. Depth maps of calibration check board at distance of 4.0 m: (a) raw depth map in front view, (b) corrected depth map in front view, (c) raw depth map in side view, (d) corrected depth map in side view.

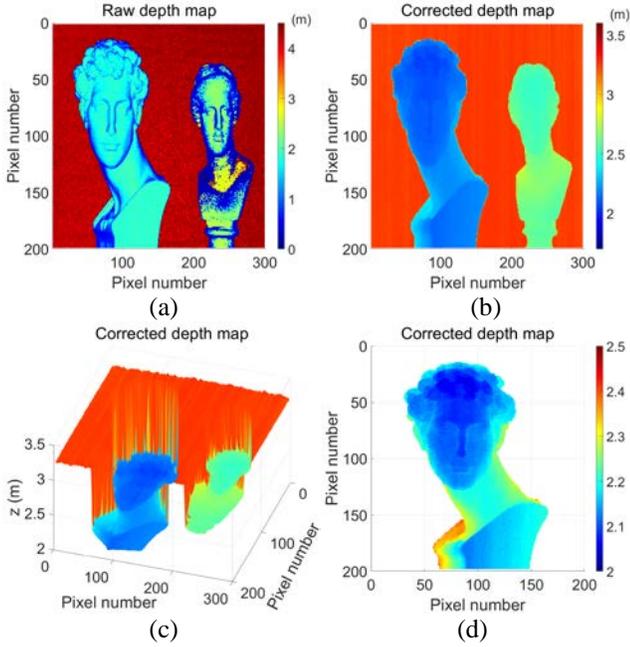

Figure 11. Depth maps of multi-objects: (a) raw depth map in front view, (b) corrected depth map in front view, (c) corrected depth map in side view, (d) zoomed Julien bust of corrected depth map.

## B. Experimental Validation of Internal Stray Light Calibration Method

The actual calibration checkboard images measured at four different distances of 1.75 m, 2.3 m, 3.0 m, and 4.0 m were used as input images in Fig. 2. After the final optimization process, the optimal internal stray light parameters were utilized to correct the depth error of each checkboard image. The measurement conditions are as follows: 20 mW of laser power, 16 μsec of integration time, 31.25 MHz single modulation frequency, bright indoor room (300 lx). The used AMCW coaxial scanning LiDAR in this paper has an optical layout same as Fig. 1 and scans the laser signal through a two-axis fast galvo scanner [6]. All the results are shown in Fig. 7 to Fig. 10.

According to Fig. 7(a), the raw depth map with resolution of 100·100, has two main clusters due to the internal stray light as explained in previous Sections. This tendency can be more easily identified in Fig. 7(c). The depth distortion due to internal stray light is corrected as Fig. 7(b) and Fig. 7(d). The standard deviation of raw depth map in Fig. 7(a) is about 0.4474 m due to the abrupt depth variation. This large deviation is reduced to 0.0142 m as shown in Fig. 7. For Fig. 8, 9, and 10, the qualitative backgrounds are the same as those of Fig. 7. The standard deviations of raw depth map in Fig. 8, 9, and 10 are 0.7730 m, 1.9248 m, and 2.2190 m, respectively. After the optimization in Fig. 2, these original standard deviations are reduced to 0.0135 m, 0.0441 m, and 0.0755 m, respectively. According to these results, as the measured distance is increased, the depth error due to internal stray light also increases. This tendency is mainly attributed to the SNR of directly reflected light to internal stray light which is lowered as the measured distance is increased. Meanwhile, as shown in Fig. 8(b) and (d), the random deviation pattern in dark region is much larger than that in relatively bright region. Such difference in deviation can be explained by the property of AMCW LiDAR. Generally, the depth deviation of AMCW LiDAR is proportional to the inverse of the amplitude of received light. Since the checkboard has a repetitive black and white pattern, the standard deviation of depth map is also inevitably changed along the image pixel position.

To validate the versatility of the proposed internal stray light calibration method, some sculptures were measured in 300·200 resolution as shown as Fig. 11. According to Fig. 11(a), (b), and (c), the original raw depth map which is enormously distorted can be corrected, restoring the original geometry of objects. To analyze the depth variation in detail, the Julien bust was zoomed as shown as Fig. 11(d). The depth map in Fig. 11(d) shows naturally smooth depth gradients in every image pixel compared to Fig. 11(a).

In summary, the depth error correction performance of the proposed calibration algorithm was validated in terms of depth deviation and versatility using checkboard images and sculptures. With proper input images of checkboard in wide ranges, the proposed stray light calibration method is anticipated to maintain accurate error correction results even in other types of AMCW coaxial scanning LiDAR.

## V. CONCLUSION

In this paper, a novel internal stray light calibration method based on GMM and PSO is proposed and demonstrated. Owing to the inherent distribution of AMCW LiDAR data, the GMM can be properly used to segment the raw amplitude map of the calibration checkboard. Using the clustered map, the depth loss is calculated and then minimized by PSO algorithm to find the optimal internal stray light parameters. The raw depth map is then corrected based on the estimated stray light information. All these processes are conducted with single modulation frequency data. According to the validation results, the average loss of depth discrepancy induced by stray light can be reduced to 3.2 mm. Using the proposed stray light calibration method, the raw depth map of geometrically complex objects could be restored based on the estimated stray light parameters. The proposed calibration algorithm in this paper can be utilized in various AMCW coaxial scanning LiDAR in that the proposed method is not dependent on the systematic information of LiDAR.